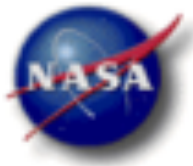
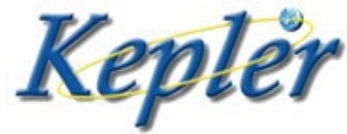

# *Uniform Modeling of KOIs*

## MCMC Data Release Notes

**NASA Ames Research Center**
**Moffett Field, CA  94035**

Prepared by: 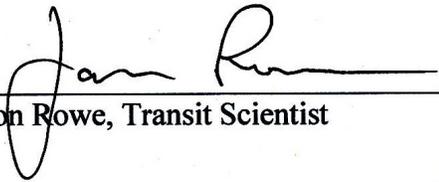 Date 3/02/15
Jason Rowe, Transit Scientist

Approved by: 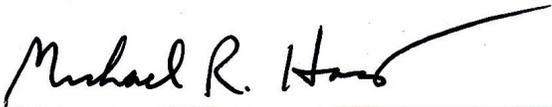 Date 3/02/15
Michael R. Haas, Science Office Director

Approved by: 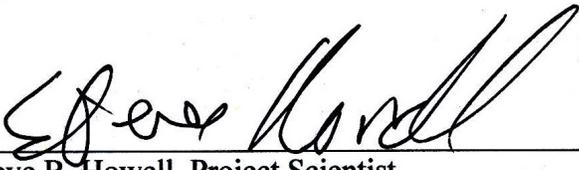 Date 3/02/15
Steve B. Howell, Project Scientist

## Document Control

Ownership

This document is part of the *Kepler* Project Documentation that is controlled by the *Kepler* Project Office, NASA/Ames Research Center, Moffett Field, California.

Control Level

This document will be controlled under KPO @ Ames Configuration Management system. Changes to this document **shall** be controlled.

Physical Location

The physical location of this document will be in the KPO @ Ames Data Center.

Distribution Requests

To be placed on the distribution list for additional revisions of this document, please address your request to the *Kepler* Science Office:

> Michael R. Haas
> *Kepler* Science Office Director
> MS 244-30
> NASA Ames Research Center
> Moffett Field, CA 94035-1000
>
> or
>
> Michael.R.Haas@nasa.gov

# DOCUMENT CHANGE LOG

| CHANGE DATE | PAGES AFFECTED | CHANGES/NOTES |
|---|---|---|
| March 2, 2015 | all | First issue |
|  |  |  |
|  |  |  |
|  |  |  |
|  |  |  |
|  |  |  |
|  |  |  |
|  |  |  |
|  |  |  |
|  |  |  |
|  |  |  |

# Table of Contents

## 1.    Introduction

This document describes data products related to the reported planetary parameters and uncertainties for the *Kepler* Objects of Interest (KOIs) based on a Markov-Chain-Monte-Carlo (MCMC) analysis.  Reported parameters, uncertainties and data products can be found at the NASA Exoplanet Archive[1]. The relevant paper for details of the calculations is Rowe et al. 2015.

The *Kepler* Mission (Borucki et al. 2010) used a 0.95-m aperture space-based telescope to continuously observe more than 150 000 stars for 4 years (Koch et al. 2010).   We model and analyze most KOIs listed at the Exoplanet Archive using the *Kepler* data. KOIs are not modeled with MCMC when the transit event does not have sufficient S/N for proper modeling (S/N >~ 7) or the KOI does not correspond to a transit event (e.g., KOI-54).  As shown in Figure 1, the planet candidates have orbital periods ranging from less than a day to greater than a year and radii ranging from that of the moon to larger than Jupiter.  These parameters are inferred by fitting a model (Mandel & Agol 2002) to the photometric time-series produced from *Kepler* observations and convolving those parameters with measured stellar parameters.  The KOI models assume that all transit-like events are produced by planets.  This assumption means that fitted parameters for stellar binaries and stellar blends will have significant systematic errors.

When a planet is observed to transit a star we see a drop in the observed flux that is proportional to the ratio of projected surface areas of the planet and star.   The duration of the transit is dictated by the orbital motion of the planet and the shape is sculpted by the tilt of the orbital plane relative to the observer and the brightness profile of the host star.  Using a transit model parameterized by the mean stellar density ($\rho*$), the ratio of the planet and star radii (r/R*), impact parameter (b), orbital period (P) and transit epoch (T0) we determine the depth, duration and shape of the transit.  Our adopted parameterization assumes that the combined mass of all transiting planets is much less than the mass of the host star.  We have also adopted non-interacting, circular orbits.  For eccentric orbits the planet-star separation during time of transit can be significantly different than the semi-major axis.  Thus a consequence of using circular orbits is that the model derived values of $\rho*$ will be systematically different from the true stellar values.

Details of the transit model can be found in §4 of Rowe et al. 2014.  The procedure for determining the model parameters can be summarized by the following tasks:

1.  Detrend Q1-Q17 PDC-Map *Kepler* photometry retrieved from MAST,

2.  Obtain a best fit model using a Levenberg-Marquardt routine (More et al. 1980),

3.  Inspect model fits with visual and numerical diagnostics and update problematic models,

4.  Measure transit timing variations (TTVs), incorporate TTVs in fits as needed,

---

[1] http://exoplanetarchive.ipac.caltech.edu

5. Run MCMC routines,

6. Inspect MCMC results and update problematic cases, and

7. Generate adopted model parameters and posteriors.

The details for each of these tasks can be found in §4 and §5 of Rowe et al. (2014) and (2015) respectively.   Here we describe the products produced by steps 2, 4 and 7 above.

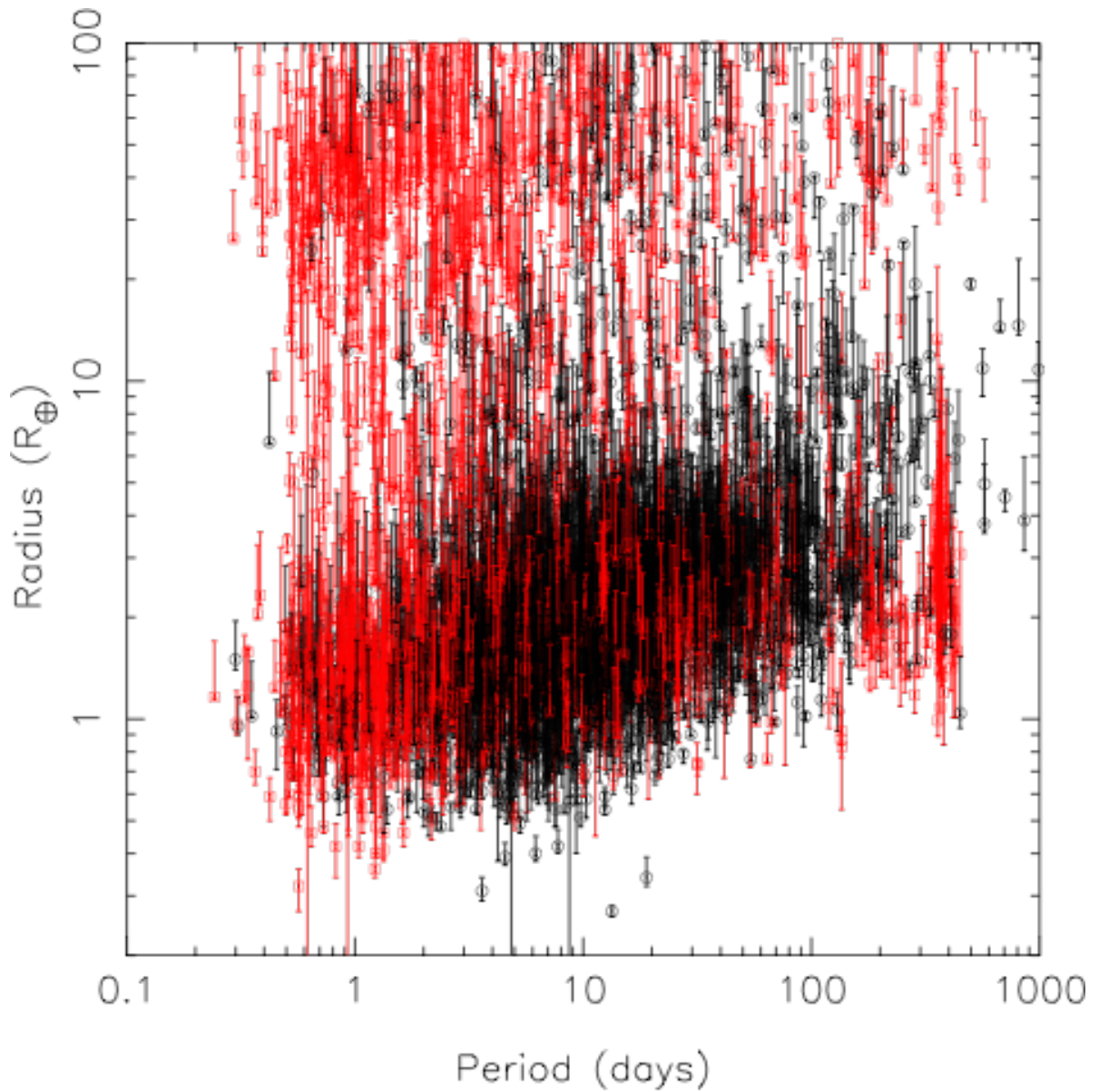

Figure 1: A plot of radius vs period for the Q1-Q12 catalogue (Rowe et al. 2015).  The black points indicate planet candidates and the red points show false-positives.

## 2.    Directory Structure

For each KOI system there is a single directory as shown in Figure 2.  The directory has the naming convention: "koiXXXX.n" where XXXX is the integer KOI number starting with 1.  All KOIs around the same star will be found in this one directory.  For example the six-planet system KOI157, with planets 157.01, 157.02, … , 157.06, is all contained in a single directory named "koi157.n".  In this directory you will find data files containing best fit parameters, transit-timing variations and Markov-Chains as indicated in Figure 2.  The next three sections explain the format of each file type.

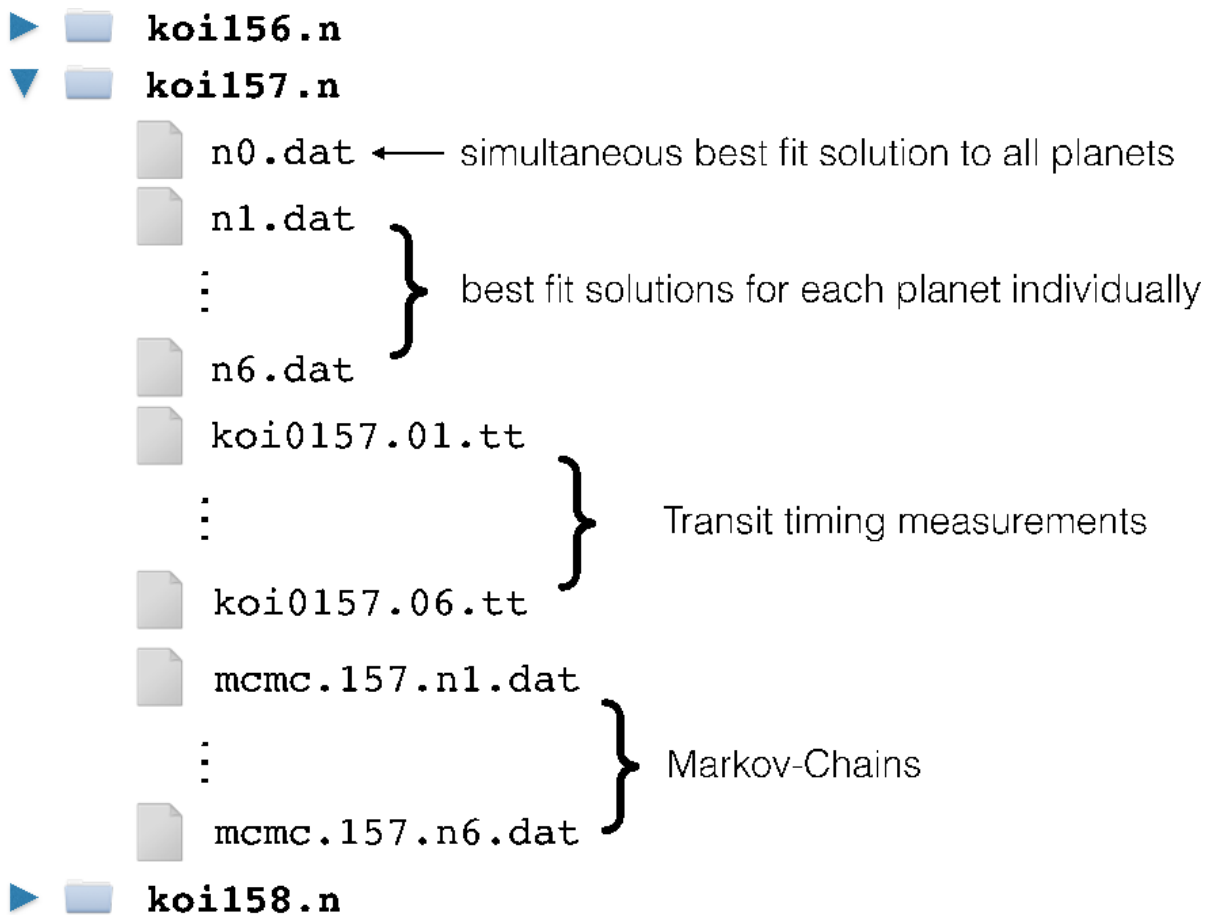

Figure 2:  Example Directory Structure

### 3.    Best Fit, Chi-Square Minimization Parameters

Each KOI system has two types of best fits: simultaneous and independent.  A file named "n0.dat" contains a global fit for all planets associated with the target star.  In the case of KOI157, the parameters are a simultaneous fit of all six known transiting planets to the *Kepler* observations.  There are also files named n1.dat, n2.dat, etc., that contain fit parameters for each planet fitted independently.  Here is an example of the contents of a best-fit file:

```
RHO   1.2236880239E+00   0.0000000000E+00  -1.0000000000E+00   0.0000000000E+00
NL1   4.2010000000E-01   0.0000000000E+00   0.0000000000E+00   0.0000000000E+00
NL2   2.5400000000E-01   0.0000000000E+00   0.0000000000E+00   0.0000000000E+00
NL3   0.0000000000E+00   0.0000000000E+00   0.0000000000E+00   0.0000000000E+00
NL4   0.0000000000E+00   0.0000000000E+00   0.0000000000E+00   0.0000000000E+00
DIL   0.0000000000E+00   0.0000000000E+00   0.0000000000E+00   0.0000000000E+00
VOF   0.0000000000E+00   0.0000000000E+00   0.0000000000E+00   0.0000000000E+00
ZPT   3.5852870878E-06   0.0000000000E+00  -1.0000000000E+00   0.0000000000E+00
EP1   7.1176453358E+01   0.0000000000E+00  -1.0000000000E+00   0.0000000000E+00
PE1   1.3024927141E+01   0.0000000000E+00  -1.0000000000E+00   0.0000000000E+00
BB1   1.3931409177E-01   0.0000000000E+00  -1.0000000000E+00   0.0000000000E+00
RD1   2.5819765742E-02   0.0000000000E+00  -1.0000000000E+00   0.0000000000E+00
EC1   0.0000000000E+00   0.0000000000E+00   0.0000000000E+00   0.0000000000E+00
ES1   0.0000000000E+00   0.0000000000E+00   0.0000000000E+00   0.0000000000E+00
KR1   0.0000000000E+00   0.0000000000E+00   0.0000000000E+00   0.0000000000E+00
TE1   0.0000000000E+00   0.0000000000E+00   0.0000000000E+00   0.0000000000E+00
EL1   0.0000000000E+00   0.0000000000E+00   0.0000000000E+00   0.0000000000E+00
AL1   0.0000000000E+00   0.0000000000E+00   0.0000000000E+00   0.0000000000E+00
```

There are 5 columns.  The first column gives the name of the fitted parameter, the second column gives the value of the fitted parameter and the fourth column determines whether a variable was fitted or held fixed.  The third and fifth columns are not used.   The fourth column is zero if the variable was held fixed during the fitting procedure, any other value indicates that the parameter was fitted.  The fitted variables are:

RHO - $\rho*$ , mean stellar density (g/cm$^3$).

NL1-4 - limb-darkening parameters.  If NL3=NL4=0, then a quadratic law was adopted, otherwise a non-linear law (Claret & Bloemen 2011) was used.

DIL - fraction of light from additional stars in the aperture that diluted the observed transit.  0 - no dilution is present, 1 - additional source corresponds to 100% of flux.

VOF - radial velocity zero point (m/s).  We did not include radial velocities in our fits.

ZPT - photometric zero point (relative).   Detrending aims to have ZPT ~ 0.

EPy - T0, time of first transit for each planet y in units of BJD-254900. For a multi-planet fit, there will be an entry for each planet: EP1, EP2, EP3,…

PEy - P, orbital period for each planet y (days).

BBy - b, impact parameter for each planet y.

RDy - r/R*, ratio of planet radius and star radius for each planet y.

ECy, ESy - eccentricity vector e cos($\omega$), e sin($\omega$) for each planet y.

KRy - radial velocity amplitude for each planet y.  Doppler beaming is included (m/s).

TEy - secondary eclipse depth for each planet y (ppm).

ELy - amplitude of ellipsoidal variations for each planet y (ppm).

ALy - amplitude of phase-curve variations from albedo for each planet y (ppm).

Table 1 gives the matching variable names as listed in the NASA Exoplanet Archive. Figure 3 shows a model fit to *Kepler* observations of *Kepler*-18b.

| Parameter | NASA Exoplanet Archive | Description |
|-----------|------------------------|-------------|
| RHO | koi_srho | fitted mean stellar density (g/cm$^3$) |
| NL1-4 | koi_ldm_coeff1,2,3,4 | limb-darkening parameters |
| ZPT | N/A | photometric zero point |
| EPy | koi_time0bk | transit epoch.  Archive uses BKJD. |
| PEy | koi_period | orbital period (days) |
| BBy | koi_impact | impact parameter |
| RDy | koi_ror | planet-star radius ratio |

Table 1: Variable names and corresponding labels from the NASA Exoplanet Archive.

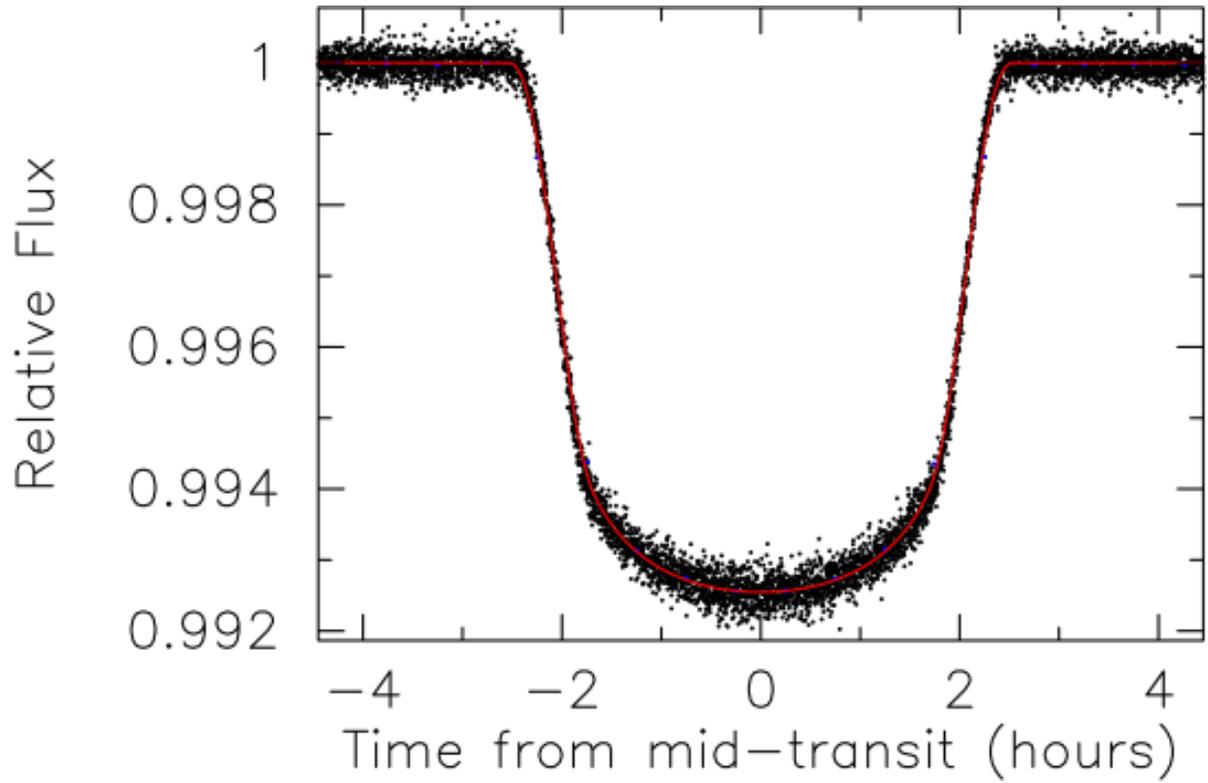

Figure 3: The transit of *Kepler*-5b (KOI-18.01). The black dots show the long-cadence photometry from *Kepler* phased to the orbital period of *Kepler*-5b. The red line shows the best-fit model.

## 4.    Transit Timing Variations

When transit-timing variations (TTVs) are detected, you will find a file with a ".tt" extension in the KOI directory.  The naming convection is "koiXXXX.0Y.tt" where "XXXX" starts at 0001 and "Y" starts at 1.  An example of the contents:

```
71.1764397960000      -4.705630182499476E-003   4.615523227458581E-003
84.2013671450000       1.198424344565296E-003   4.624050832207679E-003
97.2262944940000      -5.091152929310283E-003   5.244599663771094E-003
110.251221843000      -9.378087396754609E-003   4.027120757461007E-003
123.276149192000      -5.578341643371232E-003   4.682547373515807E-003
136.301076541000      -2.281073656746457E-003   4.438034810820665E-003
149.326003890000       1.567907907126198E-002   4.714838517840220E-003
162.350931239000       7.833819010485854E-003   5.927128525079041E-003
```

The first column is the calculated (expected) time (BJD-254900) of transit based on the orbital period of the best-fit model.  The second column gives the observed minus calculated (O-C) transit time (days) based on a fit of the transit model to the individually observed transit events.  The third column is the uncertainty on the O-C time (days).  If a ".tt" file is present, then transit-timing variations are included in the transit models.

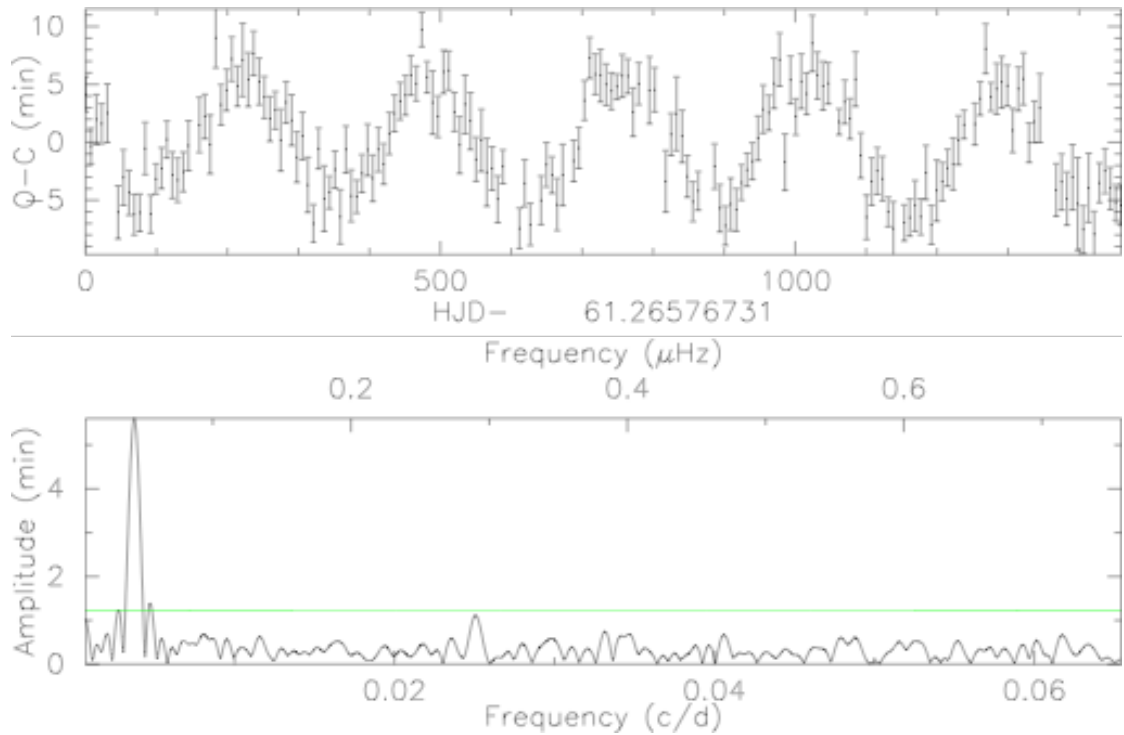

Figure 4:  The top panel shows the O-C diagram of TTVs measured from *Kepler*-18c (KOI-137.01).  The bottom panel shows a Fourier transform of the TTVs.  The green line is a 3-σ detection threshold.

## 5.    Markov-Chains

Using the best-fit model and if necessary, TTVs, a Markov-Chain-Monte-Carlo routine was run to calculate a series of model parameter sets that were then used to estimate posterior distributions for each model parameter.   The files have the naming convention, koiXXXX.nY.dat, where XXXX is the KOI number and starts at 1 and Y is the planet number starting with 1.  An example of the contents showing the first 3 lines of the file:

```
        18

  1.4914682798E+04   0 11  1.2533986362E+00   4.2010000000E-01   2.5400000000E-01
0.0000000000E+00   0.0000000000E+00   0.0000000000E+00   0.0000000000E+00
6.6148941153E-07   7.1176263245E+01   1.3024925322E+01   1.9651895472E-02
2.5704275054E-02   0.0000000000E+00   0.0000000000E+00   0.0000000000E+00
0.0000000000E+00   0.0000000000E+00   0.0000000000E+00

  1.4914718902E+04   0  1  1.2545401112E+00   4.2010000000E-01   2.5400000000E-01
0.0000000000E+00   0.0000000000E+00   0.0000000000E+00   0.0000000000E+00
6.6148941153E-07   7.1176263245E+01   1.3024925322E+01   1.9651895472E-02
2.5704275054E-02   0.0000000000E+00   0.0000000000E+00   0.0000000000E+00
0.0000000000E+00   0.0000000000E+00   0.0000000000E+00
```

The first line gives the number of parameters, Np, in the model.  For a single-planet fit, there are Np=18 parameters, for 2 planets there are Np=28, and 10 additional parameters for each additional planet.  After the first line, each subsequent line has 3+Np entries. The first column reports chi-square for the parameter set, the second column is a flag to indicate if the parameter set was accepted (0) or rejected (1), and the third column is a flag to indicate which parameter was varied (0 for a vector jump (Gregory 2011), otherwise 1 through Np).  If a chain was rejected, the last accepted chain is listed.  The model parameters are listed in order matching the contents of the best-fit file as described in §3.  Thus, the fourth column lists the mean-stellar density.

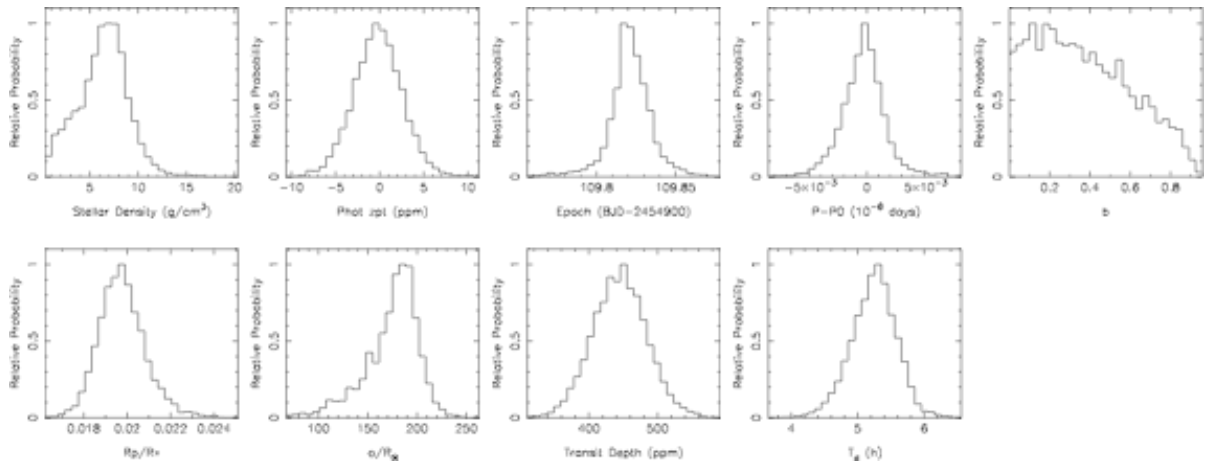

Figure 5: Posterior distributions for *Kepler*-186f (KOI-571.05).  Starting at the top and moving left we have: stellar density, photometric zero point, T0, P, b, r/R*, a/R*, transit depth and transit duration.

## 6.    Using the Files to Estimate Your Own Posteriors

MCMC routines typically require what is known as 'burn-in'.  To account for burn-in we recommend excluding at least the first ~20% of the provided chains.

To estimate any posteriors you must use both accepted and rejected flagged chains.  For example, if you wish to estimate the average value of the modeled stellar density you would calculate the average based on every entry in the fourth column of the MCMC file after excluding the first ~20%.  Figure 5 shows histograms based on a MCMC analysis of *Kepler*-186f, which can be used to estimate posterior distributions.

As another example, the values for a/R* listed at the archive were calculated via Kepler's 3rd law:

$$(a/R*)^3 = \rho* \ G \ P^2/(3 \ \pi) \ .$$

We used the values for fitted mean stellar density ($\rho*$) and period (P) to calculate a chain for the ratio a/R* for which we then estimated the median value and confidence interval.

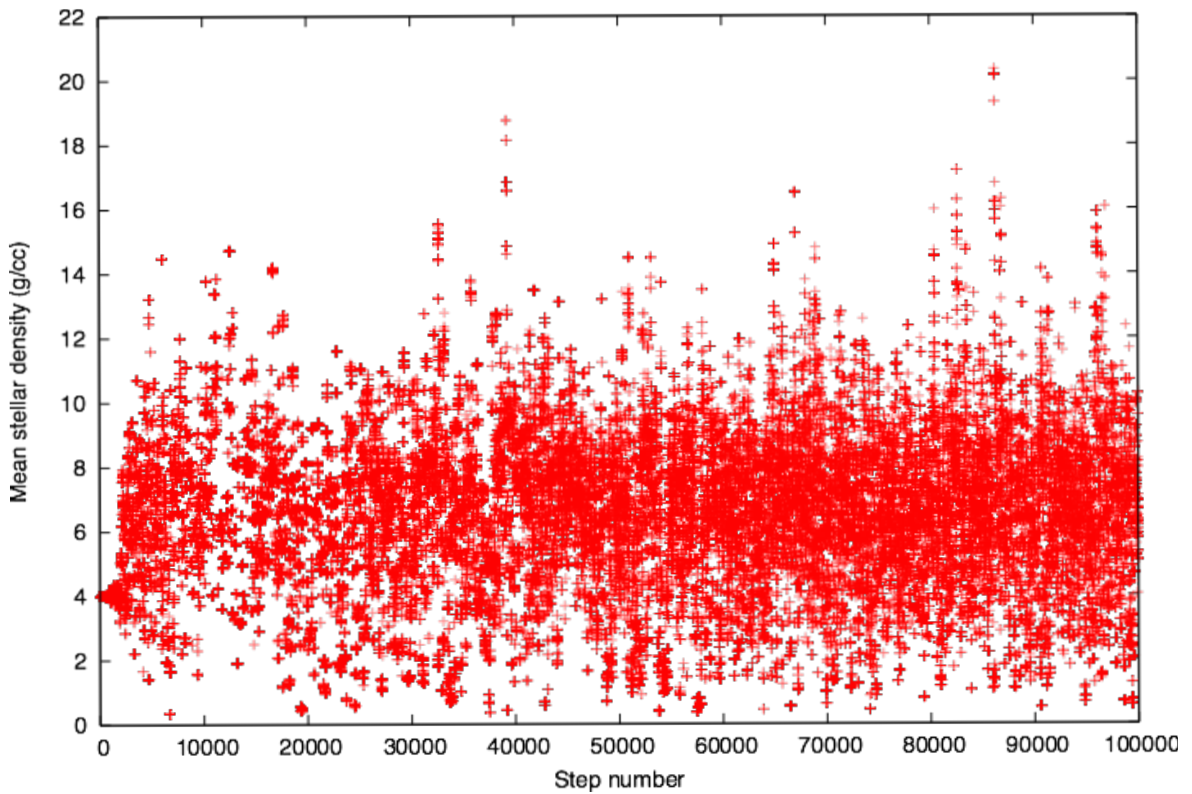

Figure 6: The value of the mean stellar density for 100 000 chains based on *Kepler* observations of *Kepler*-186f (KOI-571.05).  The first ~20% of the chains are not well mixed or behaved.  This is due to burn-in of the MCMC routine and stabilization of the Gibbs factor to achieve an acceptance rate of 20-30%.